\documentclass[letterpaper, onecolumn, 12pt]{article}

\usepackage{geometry}
\usepackage{graphicx} 
\usepackage[final]{hyperref} 
\usepackage{amsmath,amsthm} 

\usepackage{amssymb} 
\usepackage{mathtools}
\patchcmd{\proof}{\indent}{}{}{}
\usepackage{caption}
\usepackage{balance}

\newcommand{\myldots}{\kern-0.05em.\kern-0.01em.\kern-0.01em.\kern0.01em}
\newcommand{\zr}{r}
\newcommand{\hzr}{\hat{r}}
\newcommand{\R}{\mathbb{R}}
\newcommand{\I}{\mathbb{I}}
\newcommand{\0}{\mathbf{0}}
\newcommand{\norm}[1]{\left\lVert#1\right\rVert}

\newcommand\scalemath[2]{\scalebox{#1}{\mbox{\ensuremath{\displaystyle #2}}}}

\hypersetup{
	colorlinks=true,       
	linkcolor=black,        
	citecolor=black,        
	filecolor=magenta,     
	urlcolor=black         
}

\newtheorem{theorem}{Theorem}
\newtheorem{corollary}{Corollary}
\newtheorem{remark}{Remark}

\newtheorem{proposition}{Proposition}
\newtheorem{assumption}{Assumption}

\title{ \bf Configuration-Constrained Tube MPC for Tracking }

\author{Filippo Badalamenti$^1$, Sampath Kumar Mulagaleti$^2$, Alberto Bemporad$^1$,\\Boris Houska$^3$, Mario Eduardo Villanueva$^1$}
\date{\small $^1$IMT School for Advanced Studies Lucca \\[0.1cm] $^2$University of Trento \\[0.1cm] $^3$ShanghaiTech University}

\begin{document}

\maketitle
\thispagestyle{empty}

\begin{abstract}
This paper proposes a novel tube-based Model Predictive Control (MPC) framework for tracking varying setpoint references with linear systems subject to additive and multiplicative uncertainties. The MPC controllers designed using this framework exhibit recursively feasible for changing references, and robust asymptotic stability for piecewise constant references.
The framework leverages configuration-constrained polytopes to parameterize the tubes, offering flexibility to optimize their shape. The efficacy of the approach is demonstrated through two numerical examples. The first example illustrates the theoretical results, and the second uses the framework to design a lane-change controller for an autonomous vehicle.
\end{abstract}

\section{Introduction}\label{sec:introduction}
The inherent robustness of Model Predictive Control (MPC) derives from solving optimal control problems in a receding horizon manner based on state feedback, rendering it effective in managing complex control tasks.
However, in the presence of significant modelling errors or large disturbances, this may prove inadequate~\cite{Rawlings2009}, due to the mismatch between predictions and the actual behavior of the controlled process. Tube-based MPC (TMPC) has emerged as a suitable candidate to take an uncertainty model into account. The principle behind TMPC is simple: instead of optimizing over vector-valued system trajectories, one optimizes over Robust Forward Invariant Tubes (RFITs) --- together with their associated feedback policies --- that are set-valued sequences enclosing all realizations of the uncertain system.
Particular set and control parameterizations lead to different TMPC methods, for example, rigid-~\cite{MAYNE_RakovicRigidTMPC2005219,Implicit_rakovic2023}, 
homothetic~\cite{Langson2004, RAKOVIC_HomoTMPC20121631}, elastic~\cite{ETMPC_7525471, Fleming2015}, configuration-constrained tube 
MPC~\cite{villanueva2022configurationconstrained} as well as other methods based on  ellipsoidal~\cite{VILLANUEVA_ellipsoids2017311} or more general sets~\cite{Houska2019}.

In most practical applications, the control system is used to make the outputs track given set points --- typically determined by an external module, such as a steady-state optimizer~\cite{Muske1997} --- that may change over time. For example, in autonomous driving problems, the desired lateral displacement changes as the vehicle switches lanes. 
The extension of TMPC to reference tracking problems then represents a significant milestone in robust MPC design, addressing challenges arising from uncertain environments and the change in operating set points. Early TMPC approaches for reference tracking were based on a rigid tube formulation and were limited to linear systems with additive disturbances~\cite{Limon2010_tube}. Subsequent works~\cite{Hanema2017_LPV_tracking,Peschke2019} broadened this concept to include multiplicative uncertainty. All these approaches have in common that they enforce asymptotic stability by using parametric terminal sets, which requires solving optimization problems with additional variables.

\subsubsection*{Contribution}
This paper introduces a variant of Configuration-Constrained Tube MPC (CCTMPC) for varying setpoint reference tracking, based on linear prediction models subject to both additive and multiplicative uncertainty. In contrast to the CCTMPC scheme of~\cite{villanueva2022configurationconstrained} which relies on an economic MPC formulation, our approach is developed using a tracking MPC formulation. This enables the synthesis of reference tracking controllers that hinge on the solution of a single quadratic program to guarantee asymptotic stability for piecewise constant references. This extends the applicability of CCTMPC beyond regulation problems.

The use of configuration-constrained polytopes for parameterizing the RFIT results in a controller that exhibits superior properties compared to~\cite{Limon2010_tube,Hanema2017_LPV_tracking}. Notably, the need for a robust affine feedback law is eliminated, and the resulting RFITs exhibit an improved representational capability, leading to less conservativeness. A particularly relevant feature for the reference tracking problem is the unrestricted online optimization of the shape of the optimal steady-state set around the desired reference. 
The ability of our approach to handle multiplicative uncertainty, coupled with the removal of the requirement for precomputing a terminal set, renders it apt for designing tracking controllers for practical systems that can be represented by linear parameter-varying (LPV) models~\cite{Hashemi2012}. Furthermore, the scheme can seamlessly incorporate model refinement while preserving asymptotic stability, making it well-suited for the development of adaptive MPC schemes~\cite{Bujarbaruah2018}.

The paper is organized as follows: Section~\ref{sec:Sect2} introduces the problem formulation. Section~\ref{sec:Sect3} presents the tracking CCTMPC formulation as well as a novel terminal cost and related theoretical guarantees regarding recursive feasibility in the presence of reference changes, robustness, and stability, as summarized in Theorem~\ref{thm:main_stability_result} and Corollary~\ref{thm::stability}. Numerical examples are presented in Section~\ref{sec:Sect4}. Section~\ref{sec:Sect5} concludes the article.

\textit{Notation}: The symbol $\I$ represents the identity matrix. For vectors $a$ and $b$, the symbol $(a,b)$ denotes $[a^{\top} \ b^{\top}]^{\top}$.
A block-diagonal matrix formed by matrices  $Q_{1},\myldots,Q_{n}$ is denoted by $\mathrm{blkd}(Q_{1},\myldots,Q_{n})$. The symbol $\operatorname{convh}$ denotes the convex hull, and we define $\norm{z}_Q^2:=z^{\top}Qz$.

\section{Tube Model Predictive Control}\label{sec:Sect2}
This section introduces the necessary notation for our reference-tracking Tube MPC formulation. Here, we focus on Configuration-Constrained polytopic tube parameterization, as originally introduced in~\cite{villanueva2022configurationconstrained}.

\subsection{Uncertain Linear Systems}
This paper considers uncertain systems of the form
\begin{align}
    \label{eq:system}
    x^+=Ax+Bu+w \quad \text{with} \quad z=Cx+Du,
\end{align}
where $x \in \R^{n_x}$, $u \in \R^{n_u}$, and $z \in \R^{n_z}$ denote the state, input and output respectively, and $w \in \mathcal{W}$ an additive disturbance. The set $\mathcal W$ is compact and convex. While $C$ and $D$ are given matrices, $A$ and $B$ are only known 
to satisfy ${\scalemath{0.99}{(A,B)\in\Delta:=\operatorname{convh}(\{(A_1,B_1),\myldots,(A_{p},B_{p})\})}}$. 
The system is subject to closed and convex constraints $x \in \mathcal X$ and  $u \in \mathcal U$. Our aim is to design an MPC scheme that robustly steers the output $z$ towards a piecewise constant reference $\zr$ that may be updated online.

\subsection{Robust Forward Invariant Tubes}
Tube MPC methods are based on the construction of Robust Forward Invariant Tubes (RFITs). Let
\begin{align*}
\scalemath{0.95}{
    \mathcal{F}(X):= \left\{ X^+ \subseteq \R^n \ \middle| \hspace{-5pt} \begin{matrix}  & \forall \ x \in \mathcal{X}, \ \exists \ u \in \mathcal{U} \ : \vspace{2pt} \\   & Ax+Bu+w \in X^+, \vspace{2pt}  \\
    &  \hspace{5pt} \forall \ [A \ B] \in \Delta, \ w \in \mathcal{W} \end{matrix} \right\}}
\end{align*}
denote the set of one-step forward reachable sets from a given set $X \subseteq \R^{n_x}$.
A sequence $(X_t \subseteq \R^{n_x})_{t \in \mathbb{N}}$ of sets forms an RFIT if for all $t \in \mathbb{N}$, $X_{t+1} \in \mathcal{F}(X_{t})$ holds. If $X_s \in \mathcal{F}(X_s)$, then the set  $X_s$ is robust control invariant (RCI) since $X_s$ contains its forward reachable set~\cite{fb}.

\subsection{Configuration-constrained polytopic tubes}
Our focus is on polytopic RFITs defined as
\begin{align}
\label{eq:PolytopicRFIT}
    X(y_t):=\{x\in \mathbb{R}^{n_x} \, | \, Fx \leq y_t\}\;.
\end{align}
Here, $F\in\mathbb{R}^{m\times n_x}$ is a given matrix whose rows encode the facet normals 
of $X(y_t)$. We enforce \textit{configuration-constraints} 
$y_t\in\mathcal{E}:=\left\{y \in \R^m \mid Ey \leq 0\right\}$ for a given 
matrix $E$.
Such constraints fix the facial configuration of the parametric polytopes $X(y_t)$
enabling a joint hyperplane-vertex parameterization of the form 
\begin{align}
      X(y)=\operatorname{convh}(\{V_jy\in\mathbb{R}^{n_x} \, | 
\, j \in \{1,\myldots,v\} \});
\end{align}
see~\cite[Section 3.5]{villanueva2022configurationconstrained} for a construction of matrices $E$ and $V_j \in \R^{n_x \times m}$.
We define $U_j = e_j \otimes \I_{v}$, stacking the vertex control 
inputs as $u_t:=(u_{1,t},\myldots,u_{v,t})$,
such that $u_{j,t}=U_j u_t$. Let $\mathbb{S}\subseteq \mathbb{R}^{m}\times
\mathbb{R}^{vn_u}\times\mathbb{R}^{m}$ be defined as  
\begin{align}
\label{eq:S_definition}
\scalemath{0.95}{
    \mathbb{S}:=
    \left\{ (y,u,y^+) \, \middle| \,  
    \begin{aligned}
    &\forall \  (i,j) \in \{1,\myldots,p\} \times \{1,\myldots,v\}, \vspace{3pt}\\
    &F(A_iV_jy+B_iU_ju)+d \leq y^+, \vspace{3pt}\\
    & Ey \leq 0, \ V_j y \in \mathcal{X}, \ U_j u \in \mathcal{U}, 
    \end{aligned} \right\}
}
\end{align}
with $d_i:=\max\{Y_iw:w\in \mathcal{W}\}$ for all $i \in \{1,\myldots,m\}$.
\begin{proposition} 
\label{prop:RFIT_CC}
    Let $(y_t\in\mathbb{R}^{m})_{t\in\mathbb{N}}$ be such that for all $t \in \mathbb{N}$, there exists some $u_t \in \R^{vn_u}$ such that $(y_t,u_t,y_{t+1}) \in \mathbb{S}$. Then, $(X(y_t))_{t\in\mathbb{N}}$ is an RFIT.
\end{proposition}
\noindent
\textit{Proof}: The result follows from~\cite[Corollary 4]{villanueva2022configurationconstrained}.
\qed
\begin{remark}
  The method from~\cite{Loechner1997} identifies m-dimensional faces of the polytope $\big\{(x,y) \mid Fx-y\leq 0\big\}$ whose projection onto the y-coordinate space characterizes regions where the vertices of $X(y)$ depend linearly on y. Exploring its relationship with configuration constraints is a future research direction.
\end{remark}

\subsection{Optimal robust control invariant set}
We design a tube MPC scheme that computes RFITs which converge to an optimal RCI set for a given reference. 
By the definition of $\mathbb{S}$, $X(y)$ is an RCI set whenever $(y,u,y) \in \mathbb{S}$ for some $u\in\mathbb{R}^{vn_u}$. For a given reference $\zr$, we define an optimal RCI set as one whose parameters $(y,u)$ minimize a given cost $\ell(y,u,\zr)$ over $\mathbb{S}$.
As an example of such a cost, consider the function
\begin{equation*}
\ell_{1}(y,u) = \sum^{v}_{j=1} \left\Vert \left( (\overline{V} -V_j)y, \ (\overline{U} -U_j)u \right) \right\Vert^{2}_{Q_{\mathrm{v}}},
\end{equation*}
which penalizes the distance of the vertices of $X(y)$ and the vertex control inputs to their centers,
\begin{equation}
\label{eq:means_definition}
\overline{V}y = \frac{1}{v}\sum^{v}_{j=1} V_jy \quad\text{and}\quad 
\overline{U}u = \frac{1}{v}\sum^{v}_{j=1} U_ju \;.
\end{equation}
Now, let $\overline{A}$ and $\overline{B}$ be the average of matrices $A_i$ and $B_i$ over $i=1,\cdots,p$ respectively.
Assuming stabilizability of the system $x^{+} = \overline{A}x + \overline{B}u$, let the columns of $M\in\mathbb{R}^{(n_x+n_u) \times n_u}$ be a basis for the nullspace of the matrix $[\overline{A}-\I \ \overline{B}]$.  Recalling that every steady state of this system can be expressed as $(x_{\mathrm{s}\mathrm{s}},u_{\mathrm{s}\mathrm{s}}) = M\theta$, the function
\begin{equation*}
\begin{aligned}
&\ell_2(y,u,\zr,\theta) := \left\Vert
\begin{bmatrix}
\overline{V} & 0 \\
0 & \overline{U}
\end{bmatrix}
\begin{bmatrix}
y \\
u
\end{bmatrix} - M\theta
\right\Vert_{Q_{\mathrm{c}}}^2
+ 
\left\Vert \zr - 
\begin{bmatrix}C \ \ D\end{bmatrix}
M\theta \right\Vert_{Q_{\mathrm{r}}}^2
\end{aligned}
\end{equation*}
penalizes the distance of the center $(\overline{V}y,\overline{U}u)$ to the steady state $M\theta$
and the distance of its output to the reference $\zr$. We can now define a function $\ell$ given by
\begin{equation*}
\ell(y,u,\zr) := \min_{\theta} \left(\ell_1(y,u,\zr) + \ell_2(y,u,\zr,\theta)\right),
\end{equation*}
with $Q_{\mathrm{v}}$, $Q_{\mathrm{c}}$, and $Q_{\mathrm{r}}$ being positive semidefinite matrices of appropriate dimensions such that $\ell$ is a convex quadratic function. In conclusion, for a given reference $\zr$, the set $X(y_{\mathrm{o}})$ is an optimal RCI set with vertex control inputs $u_{\mathrm{o}}$, if $(y_{\rm o},u_{\rm o})$ is a minimizer of
\begin{align}
 \label{eq:OTI_set}
     \mathcal{C}_{\mathrm{o}}:=\min_{y,u} \ \ell(y,u,\zr) \ \   \text{s.t.} \ \  (y,u,y) \in \mathbb{S}.
\end{align}
\begin{assumption}
\label{ass:feasibility}
   There exists a pair $(y,u)$ satisfying $(y,u,y)\in\mathbb{S}$, such that $X(y)$ is an RCI set. 
\end{assumption}
\noindent
\textit{Problem Statement}\label{ProbStatement}: Design a CCTMPC scheme for reference tracking that guarantees $(i)$ recursive feasibility for changing references; and $(ii)$  robustness, stability, and convergence to the optimal RCI set of the current reference for piecewise constant references.

\subsection{Tube MPC with Stabilizing Initial and Terminal costs}
For a fixed reference $\zr$, one can design a CCTMPC scheme based on the optimization problem
    \begin{align}
        &\min_{\mathbf{y},\mathbf{u}} \ L_0(y_0,\zr)+ \sum_{k=0}^{N-1} \ell(y_k,u_k,\zr)+L_N(y_N,\zr) \nonumber \\ 
        &  \ \text{s.t.} \  \ F x \leq y_0, \quad y_N \in \mathbb{X}_N, \label{eq:CCTMPC_orig} \\ 
        & \qquad   (y_k,u_k,y_{k+1}) \in \mathbb{S}, \ \  \forall \  k \in \{0,\myldots,N-1\}.     \nonumber
    \end{align}
We use the bold symbols $\mathbf{y}$ and $\mathbf{u}$ to denote the vectors stacking the tube parameters $y_{0},\myldots,y_N$ and vertex inputs $u_0,\myldots,u_N$.
The state and reference measurements are $x$ and $r$, the initial and terminal costs $L_0$ and $L_N$, and the terminal region $\mathbb{X}_N$.
In~\cite{villanueva2022configurationconstrained}, a construction of $L_0$, $L_N$ and $\mathbb{X}_N$ to guarantee asymptotic stability is presented. 
In general~\eqref{eq:CCTMPC_orig} is an economic formulation as the cost, while convex, is not positive definite. 
While this formulation yields a stabilizing controller for a
\textit{fixed} reference $\zr$, adapting it to track changing references is challenging since $L_0, L_N$ and $\mathbb{X}_N$ are constructed using the solution of Problem~\eqref{eq:OTI_set} reformulated as
\begin{align}
\label{eq:OTI_multiplier}
     &\min_{y,u,y^+} \ \ell(y,u,\zr)  \ \ \text{s.t.} \ \ (y,u,y^+) \in \mathbb{S}, \ y=y^+|\lambda.
\end{align}
These ingredients depend on the primal-dual solution $(y_{\rm o}, u_{\rm o}, y_{\rm o},\lambda_{\rm o})$, which in turn depends on the reference $\zr$.
Therefore, to address the challenges highlighted in the \hyperref[ProbStatement]{\textit{Problem Statement}}, a control scheme would require one module for solving Problem~\eqref{eq:OTI_multiplier} to formulate $L_0$, $L_N$ and $\mathbb{X}_N$ whenever a change in the reference is detected, and another for solving Problem~\eqref{eq:CCTMPC_orig}. While such a scheme could operate the system safely, its cumbersome implementation can be unwieldy in practice. We address this using a reference tracking CCTMPC formulation that  eliminates the need for reformulating the underlying problem when a reference change occurs.
\begin{remark}
    Elastic Tube-MPC (ETMPC)~\cite{ETMPC_7525471} formulations can also be used to construct RFITs by optimizing $y_t$. However, such RFITs can be more conservative than those computed using CCTMPC, since CCTMPC uses vertex control laws to induce invariance instead of affine control laws. See~\cite{villanueva2022configurationconstrained} for further details.
\end{remark}
\begin{remark}
    The methods in~\cite{GUPTA2019330,mulagaleti2023parameter} etc., can be used to compute matrix $F$ of desired complexity verifying Assumption~\ref{ass:feasibility}.
\end{remark}

\section{Tube MPC for reference tracking}\label{sec:Sect3}
We now present a CCTMPC formulation that directly tackles the challenges posed in the \hyperref[ProbStatement]{\textit{Problem statement}}.
\subsection{Reference-tracking CCTMPC}
Problem~\eqref{eq:CCTMPC_orig} necessitates the use of a rotated stage cost due to presence of economic stage cost terms in the objective. Alternatively, we adopt a convex tracking cost $s(y-y_{\mathrm{s}},u-u_{\mathrm{s}})$ that vanishes at the origin and remains positive for all other points. For instance, we can define $s(z,v)=\norm{(z,v)}_Q^2$ using a positive definite matrix $Q$. Next, we formulate the optimization problem
\begin{align}
    &\hspace{-8pt}  \min_{\mathbf{y},\mathbf{u},y_{\mathrm{s}},u_{\mathrm{s}}} \ \hspace{-5pt}\scalemath{0.93}{\ell(y_{\mathrm{s}},u_{\mathrm{s}},r) + \sum_{k=0}^{N-1} 
    s(y_k,y_{\rm s},u_k,u_{\rm s})
    + \mathrm{M}(y_N,y_{\mathrm{s}},u_{\mathrm{s}})} \nonumber \\
    &\hspace{3pt}  \ \text{s.t.} \  \ F{x} \leq y_0 , \ (y_{\mathrm{s}},u_{\mathrm{s}},y_{\mathrm{s}}) \in \mathbb{S},   \ y_N \in \mathbb{T}(y_{\mathrm{s}},u_{\mathrm{s}}),     \label{eq:CCTMPC_new} \vspace{3pt}\\
     &\hspace{6pt}  \qquad   (y_k,u_k,y_{k+1}) \in \mathbb{S}, \ \  \forall \  k \in \{0,\myldots,N-1\},  \nonumber
\end{align}
\noindent
where the variables $(y_{\mathrm{s}},u_{\mathrm{s}})$ are enforced to represent an RCI set $X(y_{\mathrm{s}})$, and the tracking cost penalizes the deviation between $(y_k,u_k)$ and $(y_{\mathrm{s}},u_{\mathrm{s}})$. 
We now design a terminal cost $\mathrm{M}(y,y_{\mathrm{s}},u_{\mathrm{s}})$ and set $\mathbb{T}(y_{\mathrm{s}},u_{\mathrm{s}})$ that guarantee recursive feasibility and asymptotic stability.

\subsection{Terminal ingredients design}
As a preparation for designing the terminal cost, we define the cost-to-travel function~\cite{Villanueva2020}
\begin{align}
\label{eq:cost_to_travel}
    \mathrm{V}(y,y^+,y_{\mathrm{s}},u_{\mathrm{s}}):=&\min_{u}  \ \  \norm{ \begin{pmatrix} y - y_{\mathrm{s}}, u-u_{\mathrm{s}} \end{pmatrix}}_Q^2 \\
              &  \ \ \text{s.t.}  \ \ \ (y,u,y^+) \in \mathbb{S}, \nonumber
\end{align}
for fixed $(y_{\mathrm{s}},u_{\mathrm{s}})$.
We define $\mathrm{V}(y,y^+,y_{\mathrm{s}},u_{\mathrm{s}})=\infty$ if~\eqref{eq:cost_to_travel} is infeasible. Since the definition of $V$ is based on a tracking term, we can enforce asymptotic stability in the sequel without introducing an initial cost. Denoting
\begin{align*}
    \scalemath{0.95}{\mathcal{V}_N(\mathbf{y},y_{\mathrm{s}},u_{\mathrm{s}}):=\sum_{k=0}^{N-1} \mathrm{V}(y_k,y_{k+1},y_{\mathrm{s}},u_{\mathrm{s}}),}
\end{align*}
Problem~\eqref{eq:CCTMPC_new} can be equivalently written as%
\begin{subequations}
\label{eq:CCTMPC_CTT}
\begin{align}
    &\hspace{-8pt}  \min_{\mathbf{y},y_{\mathrm{s}},u_{\mathrm{s}}} \  \scalemath{0.99}{\ell(y_{\mathrm{s}},u_{\mathrm{s}},\hzr)+ \mathcal{V}_N(\mathbf{y},y_{\mathrm{s}},u_{\mathrm{s}})+\mathrm{M}(y_N,y_{\mathrm{s}},u_{\mathrm{s}})} \nonumber \\
    &  \text{s.t.} \ \  F\hat{x} \leq y_0, \  \ y_N \in \mathbb{T}(y_{\mathrm{s}},u_{\mathrm{s}}),  (y_{\mathrm{s}},u_{\mathrm{s}},y_{\mathrm{s}}) \in \mathbb{S}. \tag{\theparentequation} \nonumber
\end{align}
\end{subequations}
For some $\gamma \in [0,1)$, we define the terminal cost as
\begin{align}
    \mathrm{M}(y,y_{\mathrm{s}},u_{\mathrm{s}}):=&\min_{u}  \   \norm{ \begin{pmatrix} y - y_{\mathrm{s}}, u-u_{\mathrm{s}} \end{pmatrix}}_P^2 \label{eq:terminal_problem} \\
    & \hspace{5pt} \text{s.t.} \ \  (y,u,\gamma y + (1-\gamma)y_{\mathrm{s}}) \in \mathbb{S}, \nonumber
\end{align}
where $P$ is a positive definite matrix. We show that for any constant $\gamma \in [0,1)$, $M(y,y_{\mathrm{s}},u_{\mathrm{s}})$ satisfies the Lyapunov descent condition inside the implicit terminal set $\mathbb{T}(y_{\mathrm{s}},u_{\mathrm{s}}):=\big\{ y \ \mid \ \mathrm{M}(y,y_{\mathrm{s}},u_{\mathrm{s}})<\infty \big\}.$

\begin{theorem}
\label{thm:main_stability_result}
If Assumption~\ref{ass:feasibility} holds, then for any $(y_{\mathrm{s}},u_{\mathrm{s}})$ satisfying $\scalemath{0.95}{(y_{\mathrm{s}}, u_{\mathrm{s}}, y_{\mathrm{s}}) \in \mathbb{S}$, $\mathrm{M}(y,y_{\mathrm{s}},u_{\mathrm{s}})}$ is positive definite at ${y = y_{\mathrm{s}}}$. Moreover, if the matrix inequality $Q+\gamma^2 P \preceq P$ holds, then for any $\bar{y} \in \mathbb{T}(y_{\mathrm{s}},u_{\mathrm{s}})$, functions $M$ and $V$ satisfy the Lyapunov descent condition 
\begin{align}
    \label{eq:descent_condition}
\min_{y^+} \{\mathrm{M}({y}^+,y_{\mathrm{s}},u_{\mathrm{s}})+\mathrm{V}(\bar{y},{y}^+,y_{\mathrm{s}},u_{\mathrm{s}})\} \leq \mathrm{M}(\bar{y},y_{\mathrm{s}},u_{\mathrm{s}}). 
    \end{align}
\end{theorem}
\noindent
\textit{Proof}: 
Under Assumption~\ref{ass:feasibility}, there exist $(y_{\mathrm{s}},u_{\mathrm{s}})$ satisfying $(y_{\mathrm{s}},u_{\mathrm{s}},y_{\mathrm{s}}) \in \mathbb{S}$. Then, positive definiteness of $M$ holds since the objective of Problem~\eqref{eq:terminal_problem} is positive definite at $(\0,\0)$.
Regarding the descent condition, denoting the minimizer of~\eqref{eq:terminal_problem} at $y=\bar{y}$ by $\bar{u}$, we have
\begin{align}
\label{eq:rhs_part}
    \mathrm{M}(\bar{y},y_{\mathrm{s}},u_{\mathrm{s}}) = \norm{(\bar{y}-y_{\mathrm{s}}, \bar{u}-u_{\mathrm{s}})}_P^2.
\end{align}
The points $\bar{y}^+ = \gamma \bar{y} + (1-\gamma) y_{\mathrm{s}}$ and $u=\bar{u}$ are feasible for Problem~\eqref{eq:cost_to_travel} at $y^+=\bar{y}^+$. Hence,
\begin{align}
    \label{eq:lhs_P2}
    \mathrm{V}(\bar{y},\bar{y}^+,y_{\mathrm{s}},u_{\mathrm{s}}) \leq  \norm{(\bar{y}-y_{\mathrm{s}}, \bar{u}-u_{\mathrm{s}})}_Q^2.
\end{align}
If $Q+\gamma^2 P \preceq P$ holds, then~\eqref{eq:rhs_part} and~\eqref{eq:lhs_P2} along with
\begin{align}
\label{eq:lhs_P1}
   \mathrm{M}(\bar{y}^+,y_{\mathrm{s}},u_{\mathrm{s}}) \leq \gamma^2 \norm{(\bar{y}-y_{\mathrm{s}}, \bar{u}-u_{\mathrm{s}})}_P^2
\end{align}
imply~\eqref{eq:descent_condition}.
To show that~\eqref{eq:lhs_P1} holds, observe that
\begin{align}
\label{eq:main_inequalities}
    F(A_i V_j y_{\mathrm{s}}+B_iU_j u_{\mathrm{s}})+d &\leq y_{\mathrm{s}},  \\
    F(A_i V_j \bar{y}+B_iU_j \bar{u})+d &\leq \gamma \bar{y} + (1-\gamma)y_{\mathrm{s}} \nonumber
\end{align}
follow from~\eqref{eq:S_definition}.
With $\bar{u}^+ := \gamma \bar{u} + (1-\gamma) u_{\mathrm{s}}$,~\eqref{eq:main_inequalities} implies 
\begin{align}
\label{eq:yp_good}
    F(A_iV_j \bar{y}^+ + B_i U_j \bar{u}^+)+d \leq \gamma \bar{y}^+ + (1-\gamma) y_{\mathrm{s}}.
\end{align}
Since $u=\bar{u}^+$ is feasible for~\eqref{eq:terminal_problem} with $y=\bar{y}^+$,~\eqref{eq:yp_good} implies $\mathrm{M}(\bar{y}^+,y_{\mathrm{s}},u_{\mathrm{s}}) \leq \big\lVert(\bar{y}^{+}-y_{\mathrm{s}}, \bar{u}^{+}-u_{\mathrm{s}})\big\rVert_P^2$ such that $(\bar{y}^+-y_{\mathrm{s}},\bar{u}^+-u_{\mathrm{s}})=\gamma(\bar{y}-y_{\mathrm{s}},\bar{u}-u_{\mathrm{s}})$ implies~\eqref{eq:lhs_P1}.
\qed

In summary, Problem~\eqref{eq:CCTMPC_new} is implemented as
\begin{align}
    &\hspace{-2pt} \min_{\mathbf{y},\mathbf{u},y_{\mathrm{s}},u_{\mathrm{s}}} \scalemath{0.91}{{\ell}(y_{\mathrm{s}},u_{\mathrm{s}},r) + \sum_{k=0}^{N-1} \norm{ \begin{bmatrix} y_k - y_{\mathrm{s}} \\ u_k-u_{\mathrm{s}} \end{bmatrix}}_Q^2+ \norm{\begin{bmatrix} y_N-y_{\mathrm{s}} \\ u_N-u_{\mathrm{s}} \end{bmatrix}}_P^2 } 
\nonumber \\
    &  \  \ \text{s.t.} \ \  Fx \leq y_0, \ \ (y_{\mathrm{s}},u_{\mathrm{s}},y_{\mathrm{s}}) \in \mathbb{S},     \nonumber   \\
    & \qquad     \  (y_k,u_k,y_{k+1}) \in \mathbb{S}, \  \forall \  k \in \{0,\myldots,N-1\},  \nonumber   \\
    & \qquad     \ (y_N,u_N,\gamma y_N + (1-\gamma)y_{\mathrm{s}}) \in \mathbb{S}. \label{eq:CCTMPC_implemented}
\end{align}

\subsection{Closed-Loop Control Law}
Denoting the parametric minimizers of~\eqref{eq:CCTMPC_implemented} as $\mathbf{y}^*(x,r)$, $\mathbf{u}^*(x,r)$, $y_{\mathrm{s}}^*(x,r)$ and  $u_{\mathrm{s}}^*(x,r)$, the control law is constructed as a convex combination of the vertex control inputs as
\begin{align}
\label{eq:MPC_control_law}
    \mu({x},r):=\sum_{j=1}^{v} \lambda_j(x,r)u^*_{0,j}(x,r),
\end{align}
where $\lambda(x,r) \in \R^v$ solves the quadratic program
\begin{align*}
    \min_{\lambda\geq 0} \ \|\lambda\|_2^2 \ \ \text{s.t.} \ \ x=\sum_{j=1}^{v} \lambda_j V_{j} y^*_0(x,\zr), \ \sum_{j=1}^v\lambda_j=1.
\end{align*}
Then, the closed loop system associated with dynamics~\eqref{eq:system} and reference sequence $(r_t)_{t \in \mathbb{N}}$ is $x_{t+1}=A_tx_t+B_t \mu(x_t,\zr_t)+w_t,$ with output $z_t=Cx_t +D\mu(x_t,\zr_t)$.
We now present recursive feasibility and stability properties of the control scheme. We denote the optimal values of~\eqref{eq:OTI_set} and~\eqref{eq:CCTMPC_implemented} by $\mathcal{C}_{\mathrm{o}}(r)$ and $\mathcal{C}^*(x,r)$, and the minimizer of~\eqref{eq:OTI_set} by $(y_{\mathrm{o}}(r),u_{\mathrm{o}}(r))$ emphasizing the dependence of the optimal RCI set parameters on the reference $r$.
\begin{assumption}
\label{ass:strictly_convex}
    The function $\ell$ defining the optimal RCI set in Problem~\eqref{eq:OTI_set} is strictly convex.
\end{assumption}
\begin{corollary}
\label{thm::stability}
Suppose Assumption~\ref{ass:strictly_convex} holds. If $Q \succ 0$, $P \succ 0$, $\gamma \in [0,1)$ satisfy ${Q+\gamma^2 P \preceq P}$, then the following statements hold for any reference $(r_t)_{t \in \mathbb{N}}$, independent of $w_t \in \mathcal{W}$ and $[A \ B]_t\in \Delta$, $t \in \mathbb N$:

$(i)$ For the control law~\eqref{eq:MPC_control_law} with $(x,\zr)=(x_t,\zr_t)$,~\eqref{eq:CCTMPC_implemented} is feasible for all $t \in \mathbb{N}$ and $\zr_t \in \R^{n_z}$ assuming feasibility at $t=0$; 

$(ii)$ For a constant reference $\zr_t=\zr$, $\mathcal{C}^*(x_t,r) \geq \mathcal{C}_{\mathrm{o}}(r)$ and $\underset{t\to \infty}{\lim} \mathcal{C}^*(x_t,\zr)=\mathcal{C}_{\mathrm{o}}(\zr)$. Moreover, $\scalemath{0.95}{\mathcal{L}(\mathbf{y}^*(x,r),y_{\mathrm{s}}^*(x,r),u_{\mathrm{s}}^*(x,r),r):=\mathcal{C}^*(x,\zr)-\mathcal{C}_{\mathrm{o}}(\zr)}$ is a strictly descending Lyapunov function, and the minimizers converge to the optimal RCI parameters as $\scalemath{0.95}{\underset{t\to \infty}{\lim} (y_k^*(x_t,\zr),u_k^*(x_t,\zr))=} \\ \scalemath{0.95}{\underset{t\to \infty}{\lim} (y_{\mathrm{s}}^*(x_t,\zr),u_{\mathrm{s}}^*(x_t,\zr))={(y_{\mathrm{o}}(\zr),u_{\mathrm{o}}(\zr))}}$ for all $k \in \{0,\myldots,N\}$.
\end{corollary}
\noindent
\textit{Proof}:
$(i)$ Suppose Problem~\eqref{eq:CCTMPC_implemented} is feasible at some $t \in \mathbb{N}$. Denoting $\tilde{y}:=\gamma y_N^*(x_t,\zr_t)+(1-\gamma) y_{\mathrm{s}}^*(x_t,\zr_t)$ and $\tilde{u}:=\gamma u_N^*(x_t,\zr_t)+(1-\gamma) u_{\mathrm{s}}^*(x_t,\zr_t)$, Theorem~\ref{thm::stability} implies \\$\mathbf{y}=[y_1^*(x_t,\zr_t)\ \myldots\ y_{N}^*(x_t,\zr_t) \ \tilde{y}]$, $y_{\mathrm{s}}=y_{\mathrm{s}}^*(x_t,\zr_t)$, $\mathbf{u}=[u_1^*(x_t,\zr_t) \myldots u_{N}^*(x_t,\zr_t) \ \tilde{u}]$, $u_{\mathrm{s}}=u_{\mathrm{s}}^*(x_t,\zr_t)$
are feasible with $x=x_{t+1}$ and any $r=\zr_{t+1}$, since $r$ affects only the objective; 

$(ii)$ If $y_0^* \neq y_{\mathrm{s}}$, we have from~\eqref{eq:cost_to_travel} that $\mathrm{V}(y_0^*(x_t,r),y_1^*(x_t,r),y_{\mathrm{s}}(x_t,r),u_{\mathrm{s}}(x_t,r)) > 0$. Since~\eqref{eq:CCTMPC_CTT} and~\eqref{eq:CCTMPC_implemented} are equivalent, the value of~\eqref{eq:CCTMPC_CTT} with the feasible solution in ($i$) is strictly less than $\mathcal{C}^*(x_t,\zr)$. 
This implies that the tracking components of~\eqref{eq:CCTMPC_CTT} are driven to $0$, such that $\mathcal{C}^*(x_t,r)$ approaches $\mathcal{C}_{\mathrm{o}}(r)$ from above and $\mathcal{L}$ is a Lyapunov function. The minimizers converge since Assumption~\ref{ass:strictly_convex} implies~\eqref{eq:OTI_set} is strictly convex.

\begin{remark}
The study of equivalence of the Economic MPC problem~\eqref{eq:CCTMPC_orig} and the Tracking MPC problem~\eqref{eq:CCTMPC_new} could be an interesting topic for future research.
\end{remark}

\section{Numerical Examples}\label{sec:Sect4}
We present two case studies to demonstrate the efficacy of the proposed CCTMPC scheme. In both cases, the terminal cost in~\eqref{eq:terminal_problem} is computed with $\gamma=0.95$.

\subsection{Illustrative example}
\begin{figure}
    \centering
    \includegraphics[width=0.75\linewidth]{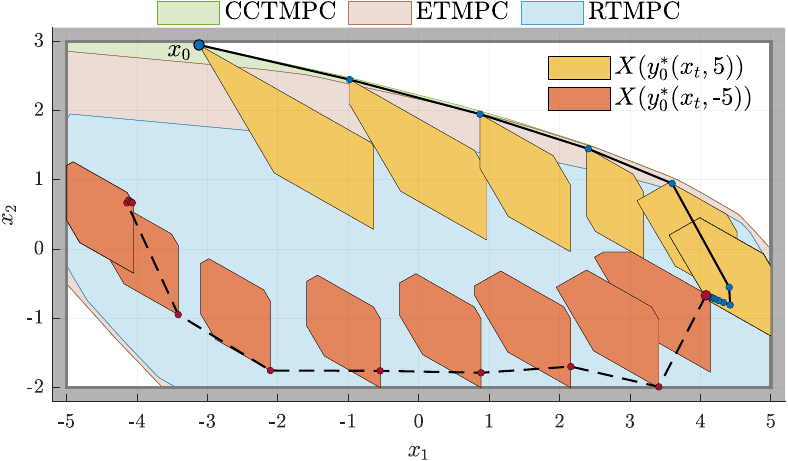}
    \captionsetup{width=\linewidth,justification=justified, singlelinecheck=false}
    \caption{Closed-loop results for System~\eqref{eq:illustrative_system}. Feasible regions of CCTMPC, RTMPC from~\cite{Limon2010_tube}, and ETMPC from~\cite{RAKOVIC2013209}, are compared. The reference is $\zr_t=5$ for $t \in [0,14]$ and $\zr_t=-5$ for $t\geq 15$. Closed-loop tubes and system trajectory are plotted for $x_0=(-3.12,2.95)$.}
    \label{fig:illustrative_example}
\end{figure}

We consider the open-loop unstable LTI system
\begin{align}
\label{eq:illustrative_system}
    x^+=\begin{bmatrix} 1.1 & 1 \\ 0 & 1 \end{bmatrix} x + \begin{bmatrix} 0.5 \\ 1 \end{bmatrix} u+w , && z=[1 \ \ 0]x,
\end{align}
with constraints $\mathcal{X}=\{x:(-5,-2)\leq x\leq (5,3)$\},  $\mathcal{U}=\{u: -1 \leq u \leq 2\}$ and $\mathcal{W}=\{w: |w|\leq (0,0.5)\}$. To design a tracking CCTMPC controller for this system, we construct matrix $F$  in~\eqref{eq:PolytopicRFIT} with $m=v=12$ as in~\cite[Remark 3]{villanueva2022configurationconstrained}, and define $\ell$ in~\eqref{eq:OTI_set} with $Q_{\mathrm{v}} = \mathrm{blkd}(10\I_2,1)$, $Q_{\mathrm{c}} = \mathrm{blkd}(\I_2,1)$ and $Q_{\mathrm{r}}=100$. We use $Q=\sum_{k=1}^{v} \mathbf{V}_k^{\top} Q_{\mathrm{v}} \mathbf{V}_k+ 10^{-3} \I_{32}$ and $P=(1-\gamma^2)^{-1}Q$, where $\mathbf{V}_k:=[\overline{V}-V_k \ \overline{U}-U_k]$, such that $Q+\gamma^2 P \preceq P$.
In Figure~\ref{fig:illustrative_example}, we plot the closed-loop results using this controller for $N=5$. In Figure~\ref{fig:Lyapunov}, the Lyapunov function $\mathcal{L}_t:=\mathcal{L}(\mathbf{y}^*(x_t,r_t),y_{\mathrm{s}}^*(x_t,r_t),u_{\mathrm{s}}^*(x_t,r_t),r_t)$ is shown.

Figure~\ref{fig:illustrative_example} compares the feasible regions of our CCTMPC, ETMPC~\cite{RAKOVIC2013209} modified for tracking as in~\cite{Limon2010_tube}, and tracking RTMPC~\cite{Limon2010_tube} controllers.
In order to parameterize the RFITs, the ETMPC controller uses the set $X(y)$, and the RTMPC uses the fixed RCI set $X(y_{\mathrm{o}}(0))$. ETMPC induces invariance in the minimal parameterized RPI set with the LQR controller with $Q=\I_2$ and $R=1$, and both schemes use the maximal positive invariant set~\cite{fb} corresponding to their respective systems as the terminal set.
CCTMPC exhibits a larger feasible region: its Hausdorff distance to the maximal RCI set is $0.0179$, compared to $0.2816$ for ETMPC and $1.1056$ for RTMPC. 
Hence, we can control the system from states close to the boundary of the MRCI set with a short horizon of $N=5$ with our tracking CCTMPC controller.

\subsection{Lane change maneuver}
We now focus on the lateral vehicle dynamics model~\cite[Section V-C]{mulagaleti2023parameter},
\begin{align}
\label{eq:lateral_model}
\scalemath{0.95}{x^+=\left(A_{0}+v^xA_{1}+(v^x)^{-1}A_{2}\right)x+Bu+B_{w}w,}
\end{align}
with $x=(e_y,e_{\psi},\dot{y},\dot{\psi})$, $u=(\delta_{\mathrm{s}},\mu_{\mathrm{b}})$ and $z=e_y$. $e_y$ [m] is the lateral deviation, $e_{\psi}$ [rad] the angular deviation,  $\dot{y}$ [m/s] the lateral velocity, $\dot{\psi}$ [rad/s] the yaw rate, $\delta_{\mathrm{s}}$ [rad] the steering angle, and $\mu_{\mathrm{b}}$ [Nm] the braking yaw moment. The disturbance is $w \in [0,100]$, and $v^x \in [40,65]/3.6$ [m/s] is the longitudinal velocity. 

Treating~\eqref{eq:lateral_model} as an LPV system with parameters $(v^x,1/v^x)$ allows us to capture its behavior as a linear system with multiplicative uncertainty. By defining the parameter set $\mathcal{P}(v^x_t):=\{(v^x,1/v^x) : v^x \in [v^x_t, 65]/3.6\}$, 
we can derive the matrices $(A_i,B_i)$ that characterize the system's multiplicative uncertainty using the vertices of of $\hat{\mathcal{P}} \supset \mathcal{P}(40)$, where
\begin{align*}
		\hat{\mathcal{P}}\coloneqq\mathrm{convh} \scalemath{0.8}{\left\{
		\begin{bmatrix}11.111\\ 0.090 \end{bmatrix},
		\begin{bmatrix}18.056\\ 0.055 \end{bmatrix},
		\begin{bmatrix}11.111\\ 0.086 \end{bmatrix},
		\begin{bmatrix}17.217\\ 0.055 \end{bmatrix}\right\}}.
\end{align*}
System constraints are $\mathcal{U}=\{u:|u|\leq (10\pi/180, 2)\}$ and $\mathcal{X}=\{x:|x|\leq (3,4,20\pi/180,3)\}$.
Matrix $F$ in~\eqref{eq:PolytopicRFIT} is computed using the approach in~\cite[Section IV-D]{mulagaleti2023parameter}; therefore, $F=\hat{S}T^{-1}$, where $\hat{S}=\left[\mathbf{I}^4\ -\mathbf{I}^4\right]^\top$ and 
\begin{equation*}
		T = 
		\scalemath{0.8}{
		\begin{bmatrix*}[r]
    0.0712 &  -0.2667  &  0.2667  &  0.0712 \\
    2.5077 &        0  &       0  & -1.4923 \\
   -0.0718 &  -0.0024  & -0.1472  & -0.0011 \\
         0 &   1.1729 &   1.8271  &       0 \\
		\end{bmatrix*}},
\end{equation*}
resulting in $m=8$ and $v=16$. The cost $\ell$ is defined with $Q_{\mathrm{v}} = \mathrm{blkd}(10^{-3}\I_4,10^{-2}\I_2)$, $Q_{\mathrm{c}} = \mathrm{blkd}(10^2\I_4,10^{-2}\I_2)$  and $Q_{\mathrm{r}}=10$, while we set $Q=\mathrm{blkd}(10^{-3}\I_8,10^{-2}\I_{32})$ and $P=(1-\gamma^2)^{-1}Q$. 
For simulation, we simplify the ``merge into faster lane" maneuver~\cite{Schildbach2015}, where a vehicle shifts from a slow lane ($\zr=-1.5$, minimum speed $40/3.6$ [m/s]) to a fast lane ($\zr=1.5$, minimum speed $60/3.6$ [m/s]), during which the parameter $v_x$ is non-decreasing, resulting in a decreasing multiplicative uncertainty over time, as $\mathcal{P}(v^x_{t_2}) \subseteq \mathcal{P}(v^x_{t_1})$ for $t_2 \geq t_1$.
The tracking CCTMPC scheme can seamlessly account for this uncertainty reduction: by recomputing the matrices $(A_i,B_i)$ at each time step using the vertices of $\mathcal{P}(v^x_{t}) \subset \hat{\mathcal{P}}$, we still ensure stability.

\begin{figure}
    \centering
    \includegraphics[width=0.75\linewidth]{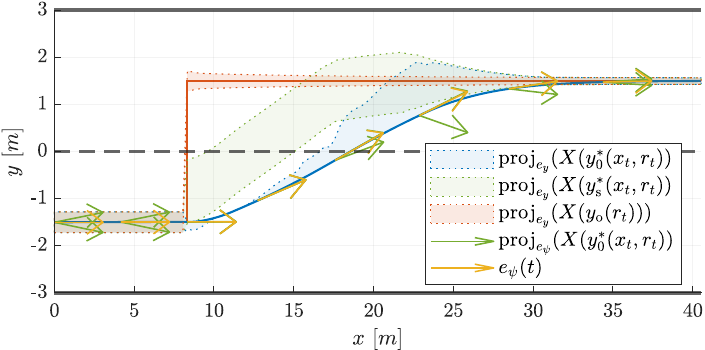}
    \captionsetup{width=\linewidth,justification=justified, singlelinecheck=false}
    \caption{Closed-loop spatial evolution of~\eqref{eq:lateral_model}.}
    \label{fig:vehicle:sim}
\end{figure}

\begin{figure}
  \centering
  \hfill
  \includegraphics[width=0.4\linewidth]{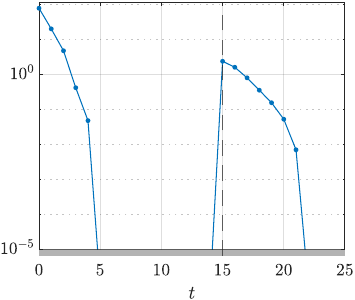}
  \hspace{0.05\linewidth}
  \includegraphics[width=0.4\linewidth]{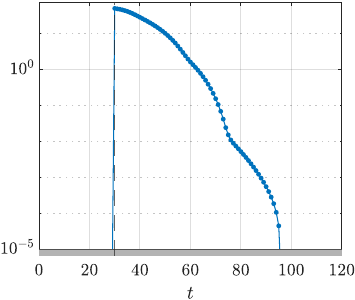}
  \hfill
  \captionsetup{width=\linewidth,justification=justified, singlelinecheck=false}
  \caption{Lyapunov function $\mathcal{L}_t$ for system~\eqref{eq:illustrative_system} (left) and \eqref{eq:lateral_model} (right) respectively. Changes in reference for the two systems are marked by vertical dashed lines.}
    \label{fig:Lyapunov}
\end{figure}

In Figure~\ref{fig:vehicle:sim} we illustrate the closed-loop dynamics using the aforementioned online model adaptation, where the CCTMPC controller is formulated with $N=3$. To visualize the set dynamics, we project $X(y_0^*(x_t,\zr_t))$ along $e_y$ (blue tube) and $e_\psi$ (angle between green arrows) states, in accordance with the spatial evolution of the system. We also show the projection of $X(y^*_{\mathrm{s}}(x_t,\zr_t))$ along $e_y$ (green tube) as well as the projection of $X(y_{\mathrm{o}}(\zr_t))$ for $e_y$ (red tube). The yellow arrow denotes $e_\psi(t)$ --- sampled every 15 time steps --- which consistently resides within the corresponding polytope projection. By explicitly accounting for reduction in multiplicative uncertainty upon moving into the fast lane, we reduce the uncertainty in the tubes as can be observed from a reduction in their width.
The Lyapunov function $\mathcal{L}_t$ is shown in Figure~\ref{fig:Lyapunov}.


\subsection{Computational aspects}
The CCTMPC scheme has $(N+1)\ v [m(1+p)+n_\mathcal{X}+n_\mathcal{U}]+m$ inequality constraints and $(N+1)(m+vn_u)$ optimization variables, where $n_{\mathcal{X}}$ and $n_{\mathcal{U}}$ are the number of halfspaces defining polytopes $\mathcal{X}$ and $\mathcal{U}$. As highlighted in~\cite[Remark 2]{villanueva2022configurationconstrained}, the matrix $E$ is typically sparse and contains redundant rows, which can be reduced using an LP solver. Trade-off between complexity of the RFIT (hence reduction in conservativeness over prediction steps) and computational burden must be taken into account in real world applications, which can be mitigated by means of high-performance QP solvers~\cite{HPIPM}. 

\section{Conclusions}
\label{sec:Sect5}
This paper has introduced a novel CCTMPC formulation for tracking piecewise constant reference signals. The MPC scheme is based on the implementation of a single convex quadratic program that simultaneously computes an RFIT and a target invariant set. Moreover, Theorem~\ref{thm:main_stability_result} and Corollary~\ref{thm::stability} have established the fact that the RFIT converges to the optimal RCI set for each constant reference, while maintaining recursive feasibility and stability guarantees in the presence of changing references. Finally, our numerical results have not only illustrated the practical applicability of the proposed CCTMPC scheme, but also its benefits compared to existing rigid-tube MPC methods for reference tracking.

\bibliographystyle{plain}
\bibliography{cctmpc_t}

\begin{thebibliography}{10}

\bibitem{fb}
F.~Blanchini and S.~Miani.
\newblock {\em Set-Theoretic Methods in Control}.
\newblock Birkhäuser, Boston, MA, 2015.

\bibitem{Bujarbaruah2018}
M.~{Bujarbaruah}, X.~{Zhang}, H.E. {Tseng}, and F.~{Borrelli}.
\newblock {{Adaptive MPC for Autonomous Lane Keeping}}.
\newblock {\em arXiv e-prints}, page arXiv:1806.04335, June 2018.

\bibitem{Fleming2015}
J.~Fleming, B.~Kouvaritakis, and M.~Cannon.
\newblock {Robust Tube MPC for Linear Systems With Multiplicative Uncertainty}.
\newblock {\em IEEE Trans. Autom. Control}, 60(4):1087--1092, 2015.

\bibitem{HPIPM}
G.~Frison and M.~Diehl.
\newblock {HPIPM: a high-performance quadratic programming framework for model predictive control}.
\newblock {\em IFAC-Pap.}, 53(2):6563--6569, 2020.
\newblock 21st IFAC World Congr.

\bibitem{GUPTA2019330}
A.~Gupta, H.~Köroğlu, and P.~Falcone.
\newblock Computation of low-complexity control-invariant sets for systems with uncertain parameter dependence.
\newblock {\em Automatica}, 101:330--337, 2019.

\bibitem{Hanema2017_LPV_tracking}
J.~Hanema, M.~Lazar, and R.~Tóth.
\newblock {Tube-based LPV constant output reference tracking MPC with error bound}.
\newblock {\em IFAC-Pap.}, 50(1):8612--8617, 2017.
\newblock 20th IFAC World Congr.

\bibitem{Hashemi2012}
S.M. Hashemi, H.~S. Abbas, and H.~Werner.
\newblock Low-complexity linear parameter-varying modeling and control of a robotic manipulator.
\newblock {\em Control Eng. Pract.}, 20(3):248--257, 2012.

\bibitem{Houska2019}
B.~Houska and M.~E. Villanueva.
\newblock {Robust optimization for MPC}.
\newblock In S.~V. Raković and W.~Levine, editors, {\em Handbook of Model Predictive Control}, pages 413--443. Birkhäuser, Cham, 2019.

\bibitem{Langson2004}
W.~Langson, I.~Chryssochoos, S.~V. Raković, and D.~Q. Mayne.
\newblock Robust model predictive control using tubes.
\newblock {\em Automatica}, 40(1):125--133, 2004.

\bibitem{Limon2010_tube}
D.~Limon, I.~Alvarado, T.~Alamo, and E.~F. Camacho.
\newblock {Robust tube-based MPC for tracking of constrained linear systems with additive disturbances}.
\newblock {\em J. Process Control}, 20(3):248--260, 2010.

\bibitem{Loechner1997}
V.~Loechner and D.~K. Wilde.
\newblock Parameterized polyhedra and their vertices.
\newblock {\em Int. J. Parallel Program.}, 25(6):525–549, 1997.

\bibitem{MAYNE_RakovicRigidTMPC2005219}
D.~Q. Mayne, M.~M. Seron, and S.~V. Raković.
\newblock Robust model predictive control of constrained linear systems with bounded disturbances.
\newblock {\em Automatica}, 41(2):219--224, 2005.

\bibitem{mulagaleti2023parameter}
S.~K. Mulagaleti, M.~Mejari, and A.~Bemporad.
\newblock {Parameter Dependent Robust Control Invariant Sets for {LPV} Systems with Bounded Parameter Variation Rate}.
\newblock {\em arXiv e-prints}, page arXiv:2309.02384, September 2023.

\bibitem{Muske1997}
K.~R. Muske.
\newblock Steady-state target optimization in linear model predictive control.
\newblock In {\em Proc. 1997 Am. Control Conf. (Cat. No.97CH36041)}, volume~6, pages 3597--3601 vol.6, 1997.

\bibitem{Peschke2019}
T.~Peschke and D.~Görges.
\newblock {Robust Tube-Based Tracking MPC for Linear Systems with Multiplicative Uncertainty}.
\newblock In {\em 2019 IEEE 58th Conf. Decis. Control (CDC)}, pages 457--462, 2019.

\bibitem{Implicit_rakovic2023}
S.~V. Raković.
\newblock The implicit rigid tube model predictive control.
\newblock {\em Automatica}, 157:111234, 2023.

\bibitem{RAKOVIC2013209}
S.~V. Raković, B.~Kouvaritakis, and M.~Cannon.
\newblock Equi-normalization and exact scaling dynamics in homothetic tube model predictive control.
\newblock {\em Syst. Control Lett.}, 62(2):209--217, 2013.

\bibitem{RAKOVIC_HomoTMPC20121631}
S.~V. Raković, B.~Kouvaritakis, R.~Findeisen, and M.~Cannon.
\newblock Homothetic tube model predictive control.
\newblock {\em Automatica}, 48(8):1631--1638, 2012.

\bibitem{ETMPC_7525471}
S.~V. Raković, W.~S. Levine, and B.~Açikmese.
\newblock Elastic tube model predictive control.
\newblock In {\em 2016 Am. Control Conf. (ACC)}, pages 3594--3599, 2016.

\bibitem{Rawlings2009}
J.~B. Rawlings and D.~Q. Mayne.
\newblock {\em {Model Predictive Control: Theory and Design}}.
\newblock Madison, WI: Nob Hill Publishing, 2009.

\bibitem{Schildbach2015}
G.~Schildbach and F.~Borrelli.
\newblock Scenario model predictive control for lane change assistance on highways.
\newblock In {\em 2015 IEEE Intell. Veh. Symp. (IV)}, pages 611--616, 2015.

\bibitem{Villanueva2020}
M.~E. Villanueva, E.~De~Lazzari, M.~A. Müller, and B.~Houska.
\newblock {A set-theoretic generalization of dissipativity with applications in Tube MPC}.
\newblock {\em Automatica}, 122(109179), 2020.

\bibitem{villanueva2022configurationconstrained}
M.~E. Villanueva, M.~A. Müller, and B.~Houska.
\newblock Configuration-constrained tube {MPC}.
\newblock {\em Automatica}, 163:111543, 2024.

\bibitem{VILLANUEVA_ellipsoids2017311}
M.~E. Villanueva, R.~Quirynen, M.~Diehl, B.~Chachuat, and B.~Houska.
\newblock {Robust MPC via min–max differential inequalities}.
\newblock {\em Automatica}, 77:311--321, 2017.

\end{thebibliography}

\balance
\end{document}